\documentclass[epj]{svjour}
\usepackage{graphics,epsfig,amssymb,amsmath,latexsym,color,pifont,graphicx,psfrag,abstract,multicol}
\newcommand{\A}{A_{ij}}
\newcommand{\B}{B_{ij}}
\newcommand{\C}{C_{ij}}

\newcommand{\E}{E_{ij}}

\newcommand{\BB}{B^{(g)}_{ij}}
\newcommand{\CC}{C^{(g)}_{ij}}

\newcommand{\EE}{E^{(g)}_{ij}}
\newcommand{\DA}{\Delta_A}
\newcommand{\DT}{\Delta_T}
\newcommand{\DDA}{\Delta_A^{(g)}}
\newcommand{\DDT}{\Delta_T^{(g)}}
\newcommand{\R}{R_{ij}}
\newcommand{\RR}{R^{(g)}_{ij}}
\newcommand{\la}{\langle}
\newcommand{\ra}{\rangle}

\begin{document}

\begin{center}

\textbf{\Large Derivative-variable correlation \\[-0.1cm] reveals the structure of \\[0.1cm] dynamical networks}\\[1.cm] 
Zoran Levnaji\'c \\
\textit{Faculty of Information Studies in Novo mesto, \\ 8000 Novo mesto, Slovenia} \\[1.cm]
\end{center}

\begin{abstract}
We propose a conceptually novel method of reconstructing the topology of dynamical networks. By examining the correlation between the variable of one node and the derivative of another node, we derive a simple matrix equation yielding the network adjacency matrix. Our assumptions are the possession of time series describing the network dynamics, and the precise knowledge of the interaction functions. Our method involves a tunable parameter, allowing for the reconstruction precision to be optimized within the constraints of given dynamical data. The method is illustrated on a simple example, and the dependence of the reconstruction precision on the dynamical properties of time series is discussed. Our theory is in principle applicable to any weighted or directed network whose internal interaction functions are known.
\end{abstract}


\section{Introduction} \label{intro}

Complex systems are ubiquitous in nature. On all scales from genes to societies, various real systems are composed of many units which collectively perform complicated tasks~\cite{mikhailov}. In the recent years, the framework of complex networks became recognized as an excellent formalism for studying complex systems. By modeling units as nodes and their interactions as links~\cite{barrat}, graph analysis methods entered physics, biology, engineering and even sociology~\cite{costa}. This allowed for a variety of real and artificial complex systems to be extensively examined, typically via computational modeling~\cite{us-pre1}. Crucial aspect of a complex network is its \textit{structure}, i.e. the topology of connections among its nodes. Properties of network structure dictate its global behavior, and are key to understanding the network's functioning and potentials for its control. For phase-repulsive oscillator models, profound intertwinement between network structure and network dynamics was recently shown~\cite{me}.

Since the structure of many natural networks is only partially known, it is of central interest to develop methods for reconstructing the network topology from the available empirical information. Various experimental techniques in this directions are already in use, specially in the context of gene regulation networks~\cite{g-h-e}. The question of network reconstruction is entering new fields, such as climatology~\cite{tsonis-donges} and neuroscience~\cite{bullmore}. Recently, the topology of a social network was inferred using mobile phone data~\cite{eagle}. In the context of oscillator networks, various theoretical~\cite{us,kralemann,luce-autariello} as well as experimental~\cite{blaha} results are available. In addition, a range of mathematical results is also available~\cite{lu}. However, most interest lies in detecting the structure of real networks from experimentally measurable outputs, such as time series~\cite{shandilya,hempel,dzeroski}.

Theoretical reconstruction methods usually rely on examining  computational network models, and are generally divided into two classes. \textit{Invasive} methods involve interfering with the network dynamics via controllable perturbation, which allows for structural data to be easily extracted~\cite{us,timme}. These methods generally give very good results, specially when the usage of perturbation is not limited in strength and frequency. However, it is often unpractical or even impossible to interact with the on-going network dynamics. \textit{Non-invasive} reconstruction methods focus on investigating the observable network outputs, such as time series of the node's dynamics, or similar measurable quantities describing the internal network states~\cite{shandilya,kralemann,hempel,xia}. The relevance of the non-invasive approach is increasingly recognized, particularly due its suitability for detecting links in biological networks~\cite{hempel,xia}. However, by not needing to interfere with the network dynamics, non-invasive approaches are typically limited to studying equilibrium network behavior, which often contains only a small amount of information on the system. Alternatively, reconstruction methods also rely on other mathematical techniques, such as control theory~\cite{dongchuan}, compressive sensing~\cite{wang}, adaptive random walk~\cite{luce1}, and even treat non-equilibrium scenarios~\cite{roudi}. It is also worth noting that, in somewhat different context, there is a large interest in investigating networks induces from time series~\cite{x-l-d}, which allow for higher-order statistical analysis of data.

In this contribution, we propose a conceptually novel non-invasive network reconstruction method. Classical studies usually involve correlations among the dynamic variables, i.e. network nodes~\cite{ren}. On the other hand, typical models of network dynamics rely on first-order differential equations (in genetic interactions, it is known that one protein's concentration determines the increase or decrease of another protein's concentration~\cite{widder}). Inspired by this, we examine the correlation between \textit{the variable of one node, and the derivative of another node}. Our central assumption is the precise knowledge of the functional forms of the intra-network interactions. As we show, depending on the quantity of network information contained in the empirical data, our method can give very precise results even for time series of length comparable to network size. Apart from being non-invasive, our method is conceptually very simple and easy to numerically implement. Based on similar hypothesis, Shandilya and Timme recently proposed another reconstruction method~\cite{shandilya}. In contrast to their result, our method avoids solving the overdetermined linear system, and in principle allows for the reconstruction error to be estimated.

The paper is organized as follows: after exposing the reconstruction method in next Section, we illustrate its implementation via simple example in Section~\ref{results}. A generalized framework of our reconstruction method in presented in Section~\ref{generalized}. The discussion of our findings and conclusions are given in Section~\ref{discussion}.


\section{The Reconstruction Method}  \label{model}

We consider a complex system composed of $N$ interacting units, which we represent as a network consisting of $N$ nodes, whose links model the node interactions. Each node is assigned a dynamical state defined by the variable $x_i \equiv x_i (t)$, where $i = 1,\hdots N$. Our first assumption is the possession of the dynamical trajectories $x_i (t_m)$, which describe the network evolution over a certain time interval. The available data consists of $N$ sequences, each containing $L$ values $x_i(t_1),\hdots x_i(t_L)$. The measurements of $x_i$ are separated by the observation interval $\delta_t = t_{m+1}-t_m$, which defines the resolution of the time series (sampling frequency). The time interval $\delta_t$ is uniform and assumed smaller than the characteristic dynamical time scale.

We further assume the evolution of the node $i$ to be governed by: 
\begin{equation} \dot x_i = f (x_i) + \sum_{j=1}^N A_{ji} h (x_j)  \, , \label{eq-1} \end{equation}
where the local node dynamics is described via function $f$, and the network (inter-node) coupling by the function $h$. The network structure is encoded in the adjacency matrix $\A$, whose element $ij$ specifies the strength with which the node $i$ acts on the node $j$. The complete dynamics of the node $i$ is the cumulative effect of the local dynamics and the contributions from its neighbors that come with different strengths. Our final assumption is the exact knowledge of both interaction functions $f$ and $h$.

We seek to reconstruct the network adjacency matrix $\A$ under the named assumptions. By having the ``fingerprint'' of the network behavior (time series of the node trajectories), we attempt to reveal its structure. Many natural systems are modeled using network equations such as Eq.\,\ref{eq-1}: examples include gene regulation and neural interactions, for which the interaction functions are widely investigated. Modern experimental techniques allow for high-resolution measurements of quantities such as gene expression, although thus obtained time series are typically short.

Inter-dependence between two variables is usually quantified through correlation~\cite{ren}, while the network models usually rely on expressing the time derivative of a node as a function of other terms, as in Eq.\,\ref{eq-1}. Inspired by this, we investigate the \textit{derivative-variable correlation} between the node $i$ (variable $x_i$) and the node $j$ (derivative $\dot x_j$). We define the following matrices:
\begin{equation} \begin{array}{lllll}
{\mathbf B}  &=&  \B &=& \langle x_i \dot x_j \rangle  \; ,  \\ 
{\mathbf C}  &=&  \C &=& \langle x_i f (x_j) \rangle  \; , \\  
{\mathbf E}  &=&  \E &=& \langle x_i h (x_j) \rangle  \; .       \label{eq-2}
\end{array} \end{equation}
$\la \cdot \ra$ denotes the time-average of a dynamical quantity (i.e., the average over the recorded time-evolution) $\la r \ra = \frac{1}{L} \sum_{m=1}^L r (t_m)$. We now re-write the Eq.\,\ref{eq-1} in the matrix form:
\begin{equation}  \mathbf{A} = ( \mathbf{B} - \mathbf{C} ) \cdot \mathbf{E}^{-1} \; , \label{eq-3} \end{equation}
which is our main reconstruction equation. To obtain a more stable function and derivative estimates, we introduce a new set of time points~\cite{shandilya}:
\[ \tau_m = \frac{t_{m+1} - t_m}{2} \; , \;\; m = 1 , \hdots, L-1 \; , \]
so that $\dot x_i (\tau_m) =  [x_i (t_{m+1}) - x_i (t_m)]/\delta_t$ and accordingly $f,h\;(\tau_m) =  [f,h\;(t_{m+1}) + f,h\;(t_m)]/2$. We rely on this calculation scheme for the computational implementation of our theory in the next Section. In principle, our method is applicable to any directed or weighted network. The reconstruction is precisely correct in the limit of large dynamical data. However, since the obtainable data are not only finite, but typically very short, our method will in general yield an approximate reconstruction.

We term the reconstructed adjacency matrix $\R$ in order to discern from the original matrix $\A$, and quantify the matrix reconstruction error as follows:
\[ \DA = \sqrt{  \frac{   \sum_{ij} ( \R - \A )^2}{ \sum_{ij}  A_{ij}^2 } } \, . \]
A natural test of the obtained $\R$, is to quantify how well does it reproduce the original data $x_i (t_m)$. To achieve this, we apply the following procedure: for all network nodes, we start the dynamics from $x_i (t_1)$, and run it using the reconstructed matrix $\R$ for the time interval $\delta_t$, i.e. until the time $t_2$. Denote thus obtained values $y_i (t_2)$, re-start the run from $x_i (t_2)$ running until $t_3$, accordingly obtaining $y_i (t_3)$, and so on. The discrepancy that the time series $y_i (t_m)$ show in comparison to $x_i (t_m)$ is the simplest measure of the reconstruction precision. We name it trajectory error $\DT$ and define as follows:
\[  \DT = \frac{1}{N} \sum_i  \sqrt{ \frac{ \big\la (x_i - y_i)^2 \big\ra }{  \big\la ( x_i - \la x_i \ra )^2 \big\ra }  } \; . \]
This way we measure point-by-point exactness of the reconstructed trajectory, which quantifies how well does it conform to the actual data. As we show in what follows, two errors are in general related, meaning that small $\DT$ suggests small $\DA$.


\section{Results}  \label{results}

\begin{figure}[!hbt] \centering 
\includegraphics[width=\columnwidth]{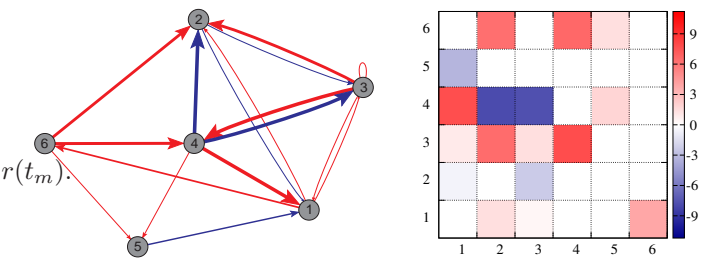} \caption{Left: graphical representation of the studied network. Link thickness illustrates the interaction strength, while color (shade) indicates positive/negative inter-node interactions. Right: network adjacency matrix.} 
\label{figure-1} \end{figure}

\begin{figure}[hbt!] \centering 
\includegraphics[width=0.9\columnwidth]{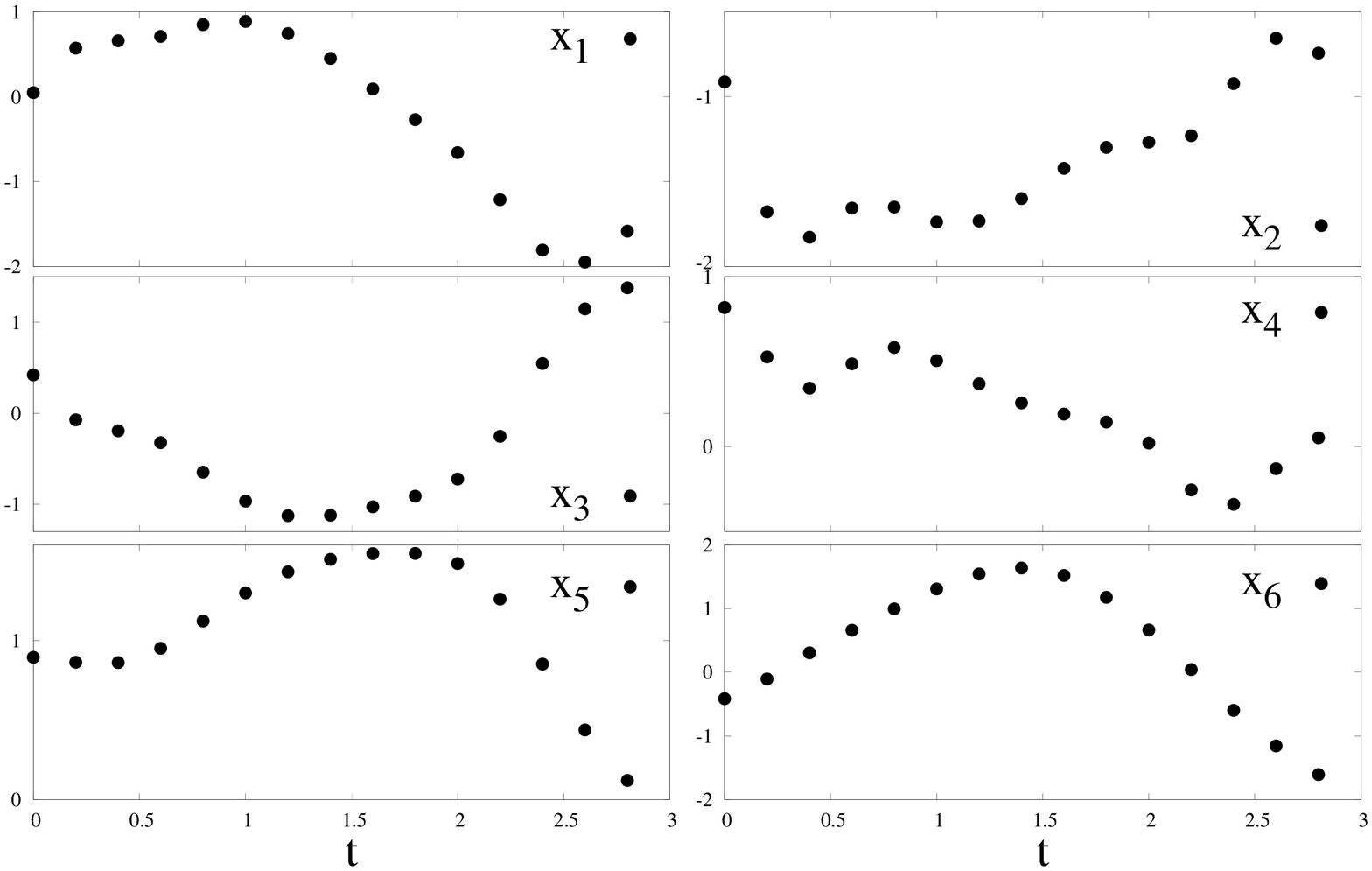} \caption{Time series for all 6 nodes for network Fig.\,\ref{figure-1}, obtained for the first set of initial conditions.} 
\label{figure-2}  \end{figure}

We test our reconstruction method using a simple illustrative example. A network with $N=6$ nodes is constructed by putting 17 directed links between randomly chosen pairs of nodes, while requiring the resulting network to be connected. Links are weighted with positive and negative weights, uniformly selected at random from $[-10,10]$. The resulting network is illustrated in Fig.\,\ref{figure-1}.
The dynamics is defined on the network via Hansel-Sompolinsky model~\cite{HS}, by taking $f = -x$ and $h = \tanh x$ in Eq.\,\ref{eq-1}. The complete dynamics on network reads: 
\begin{equation} \dot x_i = - x_i + \sum_{j=1}^{6}  A_{ji} \tanh (x_j)  \; . \label{eq-4} \end{equation}
For each node we randomly select an initial condition from $[-1,1]$, and numerically integrate Eq.\,\ref{eq-4} from time $t=0$ to $t=3$. During the run, we store 15 values for each $x_i$, equally spaced in time, starting with $x_i (t_1=0)$. The obtained time series for all nodes are shown in Fig.\,\ref{figure-2}. 

\begin{figure}[!hbt]  \centering  
\includegraphics[width=0.9\columnwidth]{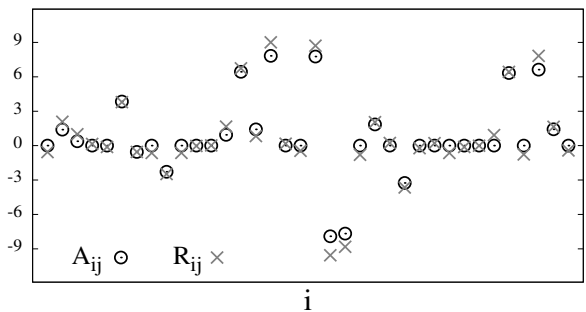}
\caption{Reconstruction via time series from Fig.\,\ref{figure-2}. Link-by-link comparison of the original values of $\A$ (circles) and the reconstructed values $\R$ (crosses). The resulting matrix and trajectory errors are $\DA=0.18$ and $\DT=0.038$, respectively.} 
\label{figure-3} \end{figure}

We now assume that these time series are obtained from an ``external'' source, such as an experimental measurement, and employ them to reconstruct the network's adjacency matrix using the method described in the previous Section. To this end, we numerically compute the derivatives $\dot x_i (\tau_m)$ and the matrices $\B$, $\C$ and $\E$, in order to obtain the reconstructed adjacency matrix $\R$ via Eq.\,\ref{eq-3}. The result is shown in Fig.\,\ref{figure-3}, through link-by-link comparison of the original and the reconstructed matrices $\A$ and $\R$.
The matrix $\R$ approximates $\A$ rather well, for both existing and non-existing links (non-zero and zero weights). The matrix error is $\DA = 0.18$, and the trajectory error is $\DT = 0.038$, indicating a good reconstruction precision.

\begin{figure}[!hbt] \centering 
\includegraphics[width=0.9\columnwidth]{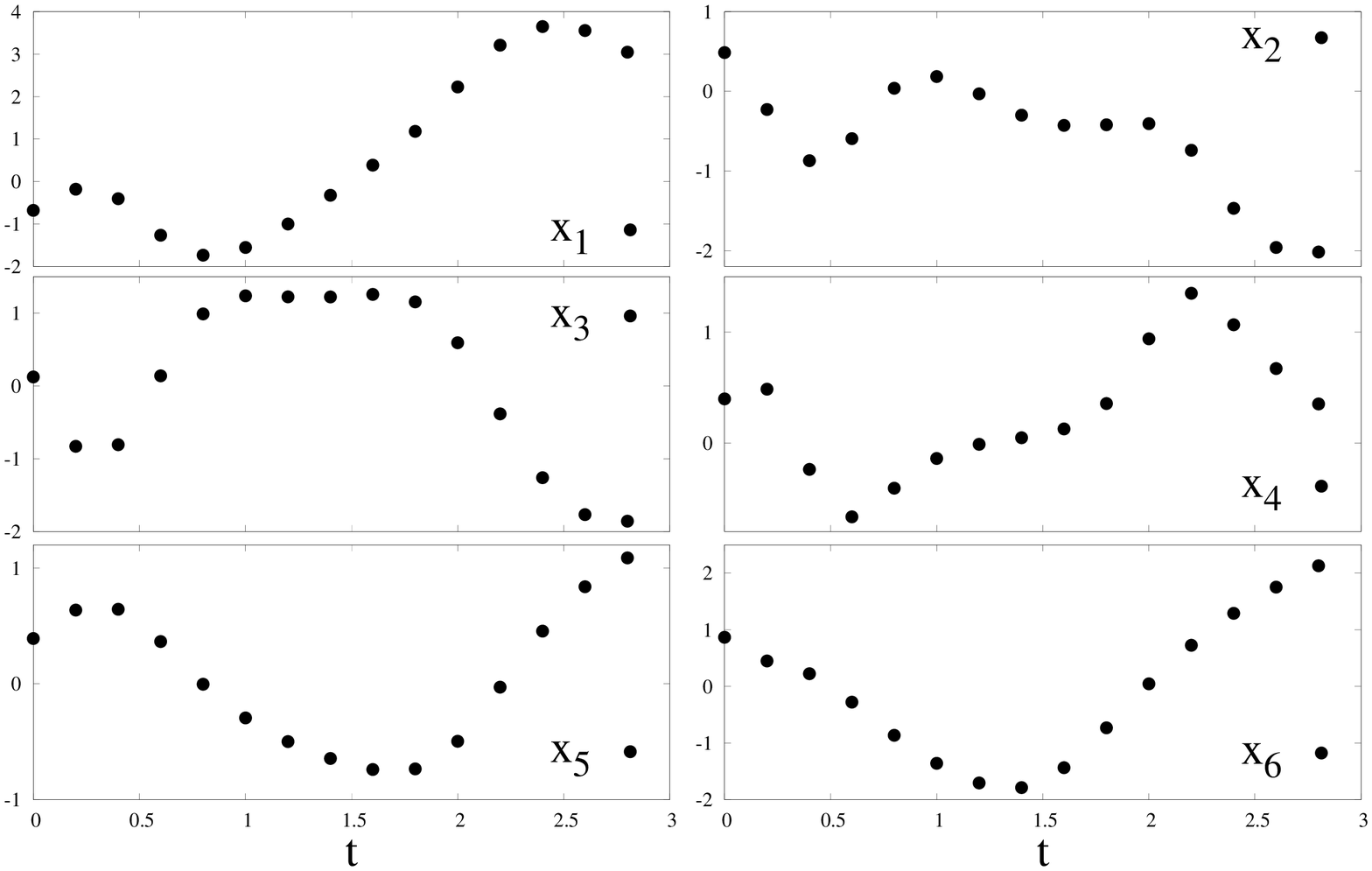}
\caption{Time series for all 6 nodes for network Fig.\,\ref{figure-1}, obtained for the second set of initial conditions.} 
\label{figure-4} \end{figure}

\begin{figure}[!hbt]  \centering  
\includegraphics[width=0.9\columnwidth]{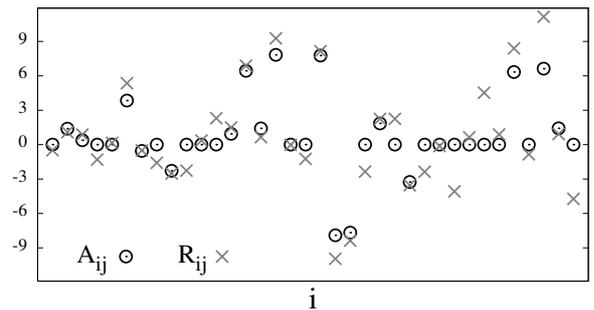}
\caption{Reconstruction via time series from Fig.\,\ref{figure-4} (done as Fig.\,\ref{figure-3} above). $\DA=0.56$, $\DT=0.050$.} 
\label{figure-5} \end{figure}

Now, we run another simulation of our dynamical system Eq.\,\ref{eq-4} with the same underlying network, but this time starting from a different set of initial conditions. The new time series of equal size and resolution is obtained and shown in Fig.\,\ref{figure-4}. We use them to reconstruct our network again, and compare two reconstructions of the same network, obtained via two sets of time series.
The new results are shown in Fig.\,\ref{figure-5}, in analogy with Fig.\,\ref{figure-3}.
The new $\R$ has the matrix error $\DA=0.56$ and the trajectory error of $\DT=0.050$, which is considerably worse than in the previous example. The reconstruction errors are bigger on most of the links, as it can be clearly seen by comparing two figures.

Despite that both time series are originating from the same dynamical network, two reconstructions are different, one of which is closer to the actual network. This shows that besides the length and resolution of the time series, the reconstruction precision crucially depends on the ``quality'' of the time series as well, i.e. on the quantity of network information contained in them. Time series showing generic patterns (e.g. periodic/synchronized oscillations) are more easy to reproduce, than the time series displaying more peculiar dynamics (e.g. strong chaos). The former dynamical data contains less extractable network information than the latter data. In fact, both considered time series are transiental in nature, but the the first ones contain more network information than the second ones, which explains the difference in precision. This is related to the concept of momentary information transfer, recently introduced in the context of coupling analysis of time series~\cite{pompe}. On the other hand, it is generally difficult to direct the experimental measurements towards maximizing the available network information. Is there a way to optimize the reconstruction precision, in the sense of extracting all usable network information, when faced with arbitrary time series? The next Section is devoted to answering this question.


\section{Generalized Reconstruction Method} \label{generalized}

The proposed reconstruction method applies to any network whose internal interactions are described by the Eq.\,\ref{eq-1}, and whose interaction functions $f$ and $h$ are known. However, the final reconstruction precision depends on a number of factors: (\textit{i}) length and resolution of the time series, also related to the precision of derivative estimates; (\textit{ii}) quantity of network information contained in the empirical data, which can be seen as the reproducibility of time series, or coverage of the dynamical phase space with data; (\textit{iii}) noise in the systems, which can manifest itself either as an error in the measurement of $x_i (t_m)$ or imprecision in the knowledge of $f$ and $h$; (\textit{iv}) invertibility of the matrix $\mathbf{E}$, which becomes unstable for $\det \mathbf{E} \approx 0$ (very important for larger networks); and finally, (\textit{v}) properties of the network itself -- some networks are more reconstructable than others, due to exhibiting qualitatively different collective dynamics. In a concrete reconstruction problem, it is difficult to isolate how much each factor contributes to $\DA$. Instead of quantifying this, we propose a generalization of our method, done towards improving and controlling the reconstruction precision, regardless of all named factors.

Our method is based on calculating the correlations between the variable $x_i$ and other terms, as defined in Eq.\,\ref{eq-2}. More generally, we can replace $x_i$ by $g(x_i)$, where $g$ is an arbitrary function, without changing the main result. Eq.\,\ref{eq-2} now becomes:
\begin{equation} \begin{array}{lllll}
\mathbf{B}^{(g)} &=&  \BB &=& \langle g(x_i) \dot x_j \rangle  \; ,  \\ 
\mathbf{C}^{(g)} &=&  \CC &=& \langle g(x_i) f (x_j) \rangle  \; , \\  
\mathbf{E}^{(g)} &=&  \EE &=& \langle g(x_i) h (x_j) \rangle  \; ,       \label{eq-5}
\end{array} \end{equation}
where notation $\mathbf{B}^{(g)}$ indicates that the matrix $\mathbf{B}$ was calculated via Eq.\,\ref{eq-5} using a pre-defined function $g$. Eq.\,\ref{eq-3}, which now becomes:
\begin{equation}  \mathbf{R}^{(g)} = \big(\mathbf{B}^{(g)}  - \mathbf{C}^{(g)} \big) \cdot \mathbf{E}^{(g)}\;^{-1} \; , \label{eq-6} \end{equation}
still holds for any choice of function $g$. As before, in the limit of very large dynamical data, the reconstruction is precisely correct for any choice of $g$. In realistic scenarios involving short time series, the reconstruction precision, except always being finite, will strongly depend on $g$. Namely, two different $g$-s will in general yield two different $\RR$-s, each with its own precision. This means that $g$ plays the role of a tunable parameter, which can be used to find the best reconstruction. Considering various choices of $g$, one can compute $\RR$ for each one of them, and define as the best that $\RR$ whose reconstructed dynamics shows minimal $\DDT$. This way, we can manipulate the reconstruction for any given time series towards finding the optimal $g$, and hence, the best $\RR$. Such $\RR$ will extract maximal available network information hidden in the time series, and improve the simple reconstruction obtained for $g(x)=x$. Moreover, the variations of $\RR$ with $g$ are related to the reconstruction precision. For a reliable reconstruction, the obtained $\RR$ will not strongly depend on changes of $g$. A bad reconstruction will be recognized by a drastic dependence of $\RR$ on $g$. Note also, that the functional properties of $g$ itself are irrelevant -- the \textit{only} role of $g$ is the computation of $\RR$.

\begin{figure}[!hbt]  \centering  
\includegraphics[width=0.9\columnwidth]{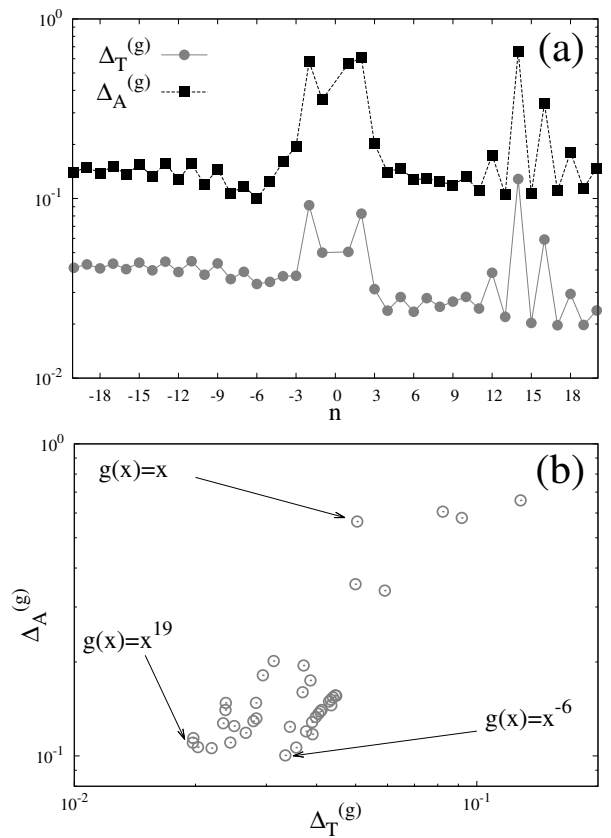}
\caption{Reconstructions from time series in Fig.\,\ref{figure-4} using 40 functions $g(x) = x^n$, for $n=-20 \hdots 20$ (except $n=0$). (a): Plot of $\DDT$ and $\DDA$ as function of the exponent $n$. (b): Scatter plot obtained by considering $\DDT$ and $\DDA$ as the two coordinates. Points corresponding to $g(x)=x$, $g(x)=x^{19}$ and $g(x)=x^{-6}$ are indicated.} 
\label{figure-6} \end{figure}

\begin{figure}[!hbt]  \centering  
\includegraphics[width=0.9\columnwidth]{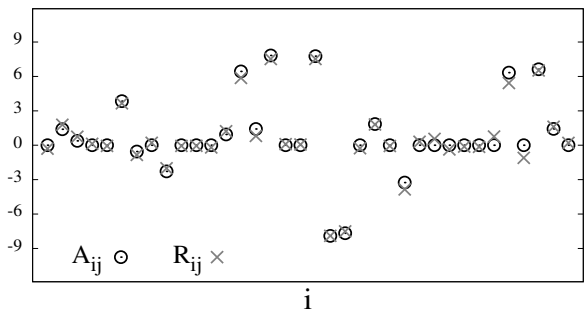}
\caption{Reconstruction via time series from Fig.\,\ref{figure-4} using the function $g(x)=x^{19}$ (analogous to Fig.\,\ref{figure-5}). Matrix and trajectory errors $\DDA=0.11$ and $\DDT=0.020$, respectively.}
\label{figure-7} \end{figure}

To illustrate the implementation of our generalized method, we examine again the second time series shown in Fig.\,\ref{figure-4} (``lower quality'' ones). For simplicity, we choose the set of functions $g(x) = x^n$, where for $n$ we take integers between -20 and 20 (except 0). The network is reconstructed using Eq.\,\ref{eq-6} for each such $g$, and the corresponding $\DDT$ and $\DDA$ are calculated. The results are reported in Fig.\,\ref{figure-6}: in (a), we show the dependence of $\DDT$ and $\DDA$ on the exponent $n$, while in (b) we show the scatter plot of $\DDT$ vs $\DDA$. 
Clearly, different choices of $g$ lead to very different reconstruction precisions. The correlation between the two errors is visible in both plots, suggesting that smaller $\DDT$, on average, leads to a smaller $\DDA$. Moreover, the dependence of $\DDT$ and $\DDA$ on the exponent $n$ appears relatively smooth. The best reconstruction, in terms of matrix error $\DDA$, is found for $g(x)=x^{-6}$ ($\DDT=0.033$, $\DDA=0.10$). However, the reconstruction displaying minimal $\DDT$ is obtained for $g(x)=x^{19}$ ($\DDT=0.020$, $\DDA=0.11$). Despite the minimal $\DDT$ not coinciding with the minimal $\DDA$, their values are still relatively close. Note that, as indicated in Fig.\,\ref{figure-6}b, both of these results are much better than what initially obtained for $g(x)=x$. In addition, the obtained precision is also better than the one found for the first (``better quality'') time series from Fig.\,\ref{figure-2}. This indicates, that despite the lower network information content in the time series from Fig.\,\ref{figure-4}, we can considerably improve the reconstruction precision by adequately tuning the function $g$. Of course, searching the $g$-functional space beyond these 40 functions is likely to yield even better precision. We show the reconstruction for $g(x)=x^{19}$ in Fig.\,\ref{figure-7}, in analogy with Fig.\,\ref{figure-3} and Fig.\,\ref{figure-5}.

The role of $g$ is to compensate for errors in the reconstruction, occurring for the reasons discussed at the beginning of this Section. As already noted, considering more (linearly independent) $g$-s will improve the precision. However, the question of selecting the optimal $g$ which extracts \textit{all} the network information contained in the time series remains open. Most straightforwardly, one can search for such $g$ via Monte Carlo method, using randomly chosen functions, or through techniques such as evolutionary optimization algorithms~\cite{ja-srep}. On the other hand, in the actual reconstruction problem we can only measure $\DDT$. From Fig.\,\ref{figure-6}, this leads to $g(x)=x^{19}$ as our ``best guess'' for $\RR$. Here, one can seek to estimate $\DDA$ by examining the changes of such $\RR$ for small variations around $g(x)=x^{19}$.


\section{Discussion and Conclusions}   \label{discussion}

We presented a novel method of reconstructing the topology of a general dynamical network from the time series and knowledge of interaction functions. By investigating the derivative-variable correlation, our method reduces to a simple matrix equation, as often studied in linear systems theory~\cite{joao}. Notably, our method works reasonably well for time series of length comparable to the network size (we reconstructed $6 \times 6$ matrix from $6 \times 15$ data points).

\begin{figure}[!hbt]  \centering  
\includegraphics[width=0.9\columnwidth]{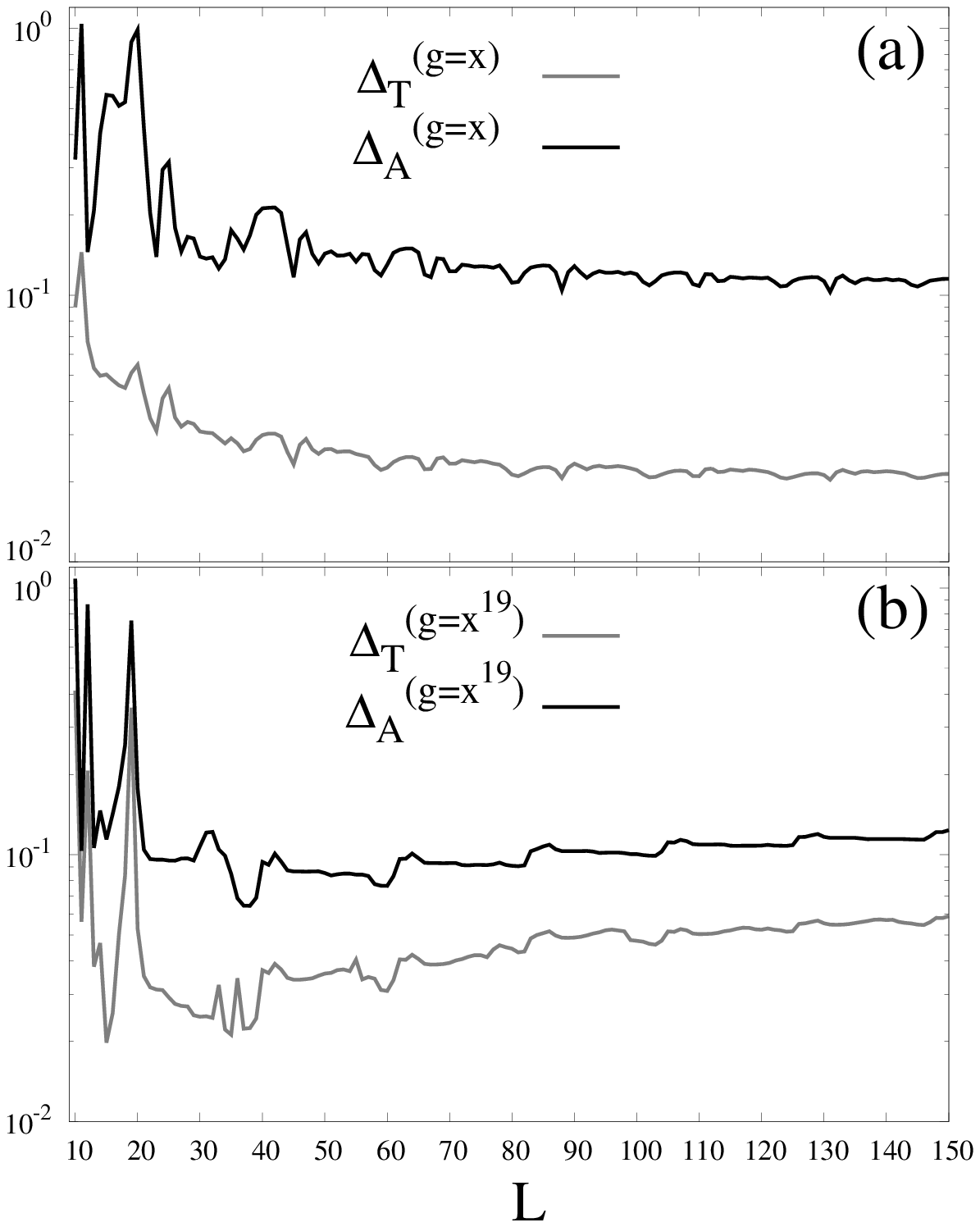}
\caption{Errors $\DDT$ and $\DDA$ as function of the time series length $L$, computed for the time series from Fig.\,\ref{figure-4}. Functions $g(x)=x$ and $g(x)=x^{19}$ are considered, as indicated in the legend.}
\label{figure-8} \end{figure}

To analyze in more detail the convergence properties of our reconstruction theory, we consider the dependence of errors $\DDT$ and $\DDA$ on the length of the time series. We consider the first 150 time points of time series from Fig.\,\ref{figure-4}. In Fig.\,\ref{figure-8}, we show the plots of $\DDT$ and $\DDA$ as function of the time series length $L$, for functions $g(x)=x$ and $g(x)=x^{19}$.
For small $L \lesssim 20$, all plots are very erratic, while for larger $L \gtrsim 50$, they become more stable, eventually saturating at given constant values. The behavior at small $L$ confirms the importance of choosing the best $g$, due to its major influence on the reconstruction precision. Interestingly, the precision does not always improve with increasing $L$. Namely, the considered time series whose initial transient is shown in Fig.\,\ref{figure-4}, eventually become periodic for all nodes. This explains the finite precisions obtained for $L \gg 1$: they are intimately related with the network information content of the time series. Note that final precisions $\DDA$ depend less on the choice of $g$. Moreover, this accounts for the reduction of precision observed for $g(x)=x^{19}$: with increase of $L$, examined time series include more and more periodic data, which carries less information than the transient data, in turn degrading the precision.

A significant limitation of our method is the ubiquity of noise. Simple estimation of derivatives is very sensitive to noise in $x_i$. However, by using the appropriate data smoothing techniques~\cite{simonoff}, one might still construct a reasonable approximation. Noise can also be present in form of stochastic time-evolution of our system (additive noise term in Eq.\,\ref{eq-1}). Here, if the noise intensity is known, one could modify Eq.\,\ref{eq-6} to include this term as well. Our strongest hypothesis is the precise knowledge of the interaction functions $f$ and $h$. Lifting this assumption would greatly enhance the generality of our theory, and render it far more applicable to the actual empirical data. However, when the approximate functional forms are known, interaction functions can be expanded in series, facilitating their reconstruction. This would mean that for each $g$, we obtain not just $\RR$, but also $f^{(g)}$ and $h^{(g)}$. This leads to a possibility of obtaining many different networks, all reproducing empirical data equally well, but in combination with different interaction functions. Another limiting factor regards the reconstruction of large real networks with $N \gg 1$. For such networks, the typical length of the obtainable time series is much smaller than the network size. Moreover, matrix inversion can become unstable for larger $N$, inducing additional imprecisions. Future developments in this direction will necessarily involve inversion check, done for instance via singular value decomposition.

The problem of network reconstruction is similar to the problem of designing a network with a prescribed dynamics. One can in principle use our method to design a network that displays given time series, by specifying the error $\DDT$, which here plays the role of tolerance. Of course, the key difference between network reconstruction and network design, is the interpretation of the solutions. In case of many different networks that satisfying $\DDT=0$, any of them is the solution of the design problem. In reconstruction theory however, one faces the issue of determining which one of them is behind the observed dynamics. We also note that our key assumption is the mathematical form of network interactions given by Eq.\,\ref{eq-1}. While a similar theory could be developed for any known form of Eq.\,\ref{eq-1}, the problem arises for networks whose interaction form is not (precisely) known, which in often the case in real (e.g. biological) networks. Our method is in principle extendable to networks with multivariate node dynamics, where $\mathbf{x}_i (t_n)$ is now a vector. This will however involve additional complications, depending on which components of the node $i$ interact with which components of the node $j$ linked to $i$. A similar situation is encountered in networks with coupling that involves two variables, such as Kuramoto model~\cite{arenas}. On the other hand, the invasive version of our theory would involve time series measured immediately after an external perturbation. Such additional assumption would greatly improve our method, by providing more transient dynamics which contains more extractable network information. Finally, we note that the key contribution of our theory is its conceptual novelty, coming from examining derivative-variable correlations. This calls for a comparative study of the reconstruction methods, primarily using experimental data. Besides identifying the best methods depending on the reconstruction context, this comparison will also potentially improve the existing and suggest entirely new methods.


\section{Acknowledgments}
This work was supported by ARRS Research Program P1-0383 ``Complex Networks''. Many thanks to Arkady for suggesting the original idea, in addition to Misha, \v Suki, Bernard and Matja\v z for constructive feedback.

\end{document}